\documentclass[pdftex, 5p, preprint]{elsarticle}

\usepackage{amsmath} 
\usepackage{amstext} 
\usepackage{amssymb}
\usepackage{url}

\def\eqref#1{(\ref{#1})}
\def\l{\langle}

\def\r{\rangle}

\begin{document}

\begin{frontmatter}
\title{Simulating spin models on GPU}
\author{Martin Weigel}
\ead{weigel@uni-mainz.de}
\address{Institut f\"ur Physik, KOMET 331, Johannes Gutenberg-Universit\"at Mainz,
  Staudinger Weg 7, 55128 Mainz, Germany}

\begin{abstract}
  Over the last couple of years it has been realized that the vast computational
  power of graphics processing units (GPUs) could be harvested for purposes other
  than the video game industry. This power, which at least nominally exceeds that of
  current CPUs by large factors, results from the relative simplicity of the GPU
  architectures as compared to CPUs, combined with a large number of parallel
  processing units on a single chip. To benefit from this setup for general computing
  purposes, the problems at hand need to be prepared in a way to profit from the
  inherent parallelism and hierarchical structure of memory accesses. In this
  contribution I discuss the performance potential for simulating spin models, such
  as the Ising model, on GPU as compared to conventional simulations on CPU.
\end{abstract}


\end{frontmatter}

\section{Introduction}

Owing to a combination of an improved toolset of simulational machinery and methods
of data analysis and the exponential increase in available computer power observed
over the past four decades, computer simulations such as the Monte Carlo method have
at least drawn level with the more traditional perturbative approaches for studying a
plethora of problems in statistical physics \cite{binder:book2}, ranging from
critical phenomena \cite{pelissetto:02} over the physics of disordered systems
\cite{young:book} to soft matter and biological problems \cite{holm:05}. This success
notwithstanding, a range of notoriously hard problems appear to create an insatiable
appetite for more powerful computational devices to finally settle a number of
long-standing questions. Among such problems are, for instance, the quest of
understanding the nature of the spin glass phase \cite{kawashima:03a} or the protein
folding problem. To achieve results beyond the reach of the available standard
computational resources of the time, there has been a tradition of designing special
purpose computers, e.g., for calculations in lattice field theory \cite{goldrian:08}
or the simulation of spin models \cite{bloete:99a,belleti:09}.

Since the design and programming of such dedicated machines regularly requires a
large effort in terms of monetary and human resources, recently scientists have
started to adopt the use of graphics processing units for general purpose
computational tasks in the hope of harvesting their nominally vast computational
power, on par with some devices based on FPGAs, without the need of time-consuming
work at and near the hardware level \cite{tomov:05,meel:07,preis:09}. By design, GPUs
are optimized for manipulating a large number of graphics primitives in parallel,
which often amounts to simple, floating-point matrix calculations. In contrast to
current CPUs, they are not designed to cope with ``unexpected'' branches in the code,
or for executing a single-threaded program as fast as possible. While this makes GPUs
not well suited as drop-in replacements for CPUs for interactive computing, their
highly parallel architecture might well be taken advantage of in scientific
calculations with an often high degree of vectorizable or parallelizable code. Their
original design for graphics calculations, however, entails certain design features
which are not necessarily optimal for scientific computational tasks, such as a
special hierarchy of memory organization or a restriction to (efficient)
floating-point calculations only in single precision arithmetics, which only has been
alleviated in the very latest generation of cards.

While the first applications of general purpose computing on GPUs were performed
directly in graphics programming languages such as OpenGL \cite{tomov:05}, access to
these devices for scientific applications has been considerably simplified with the
advent of language extensions such as NVIDIA CUDA \cite{cuda} and OpenCL \cite{opencl}
for performing general purpose computing on GPUs. The application presented here
was coded on the NVIDIA architecture using the CUDA framework, which is a high-level
extension to the C language family.

\section{Relevant features of GPU architecture\label{sec:hardware}}

\begin{figure}[tb]
  \centering
  \includegraphics[width=0.475\textwidth]{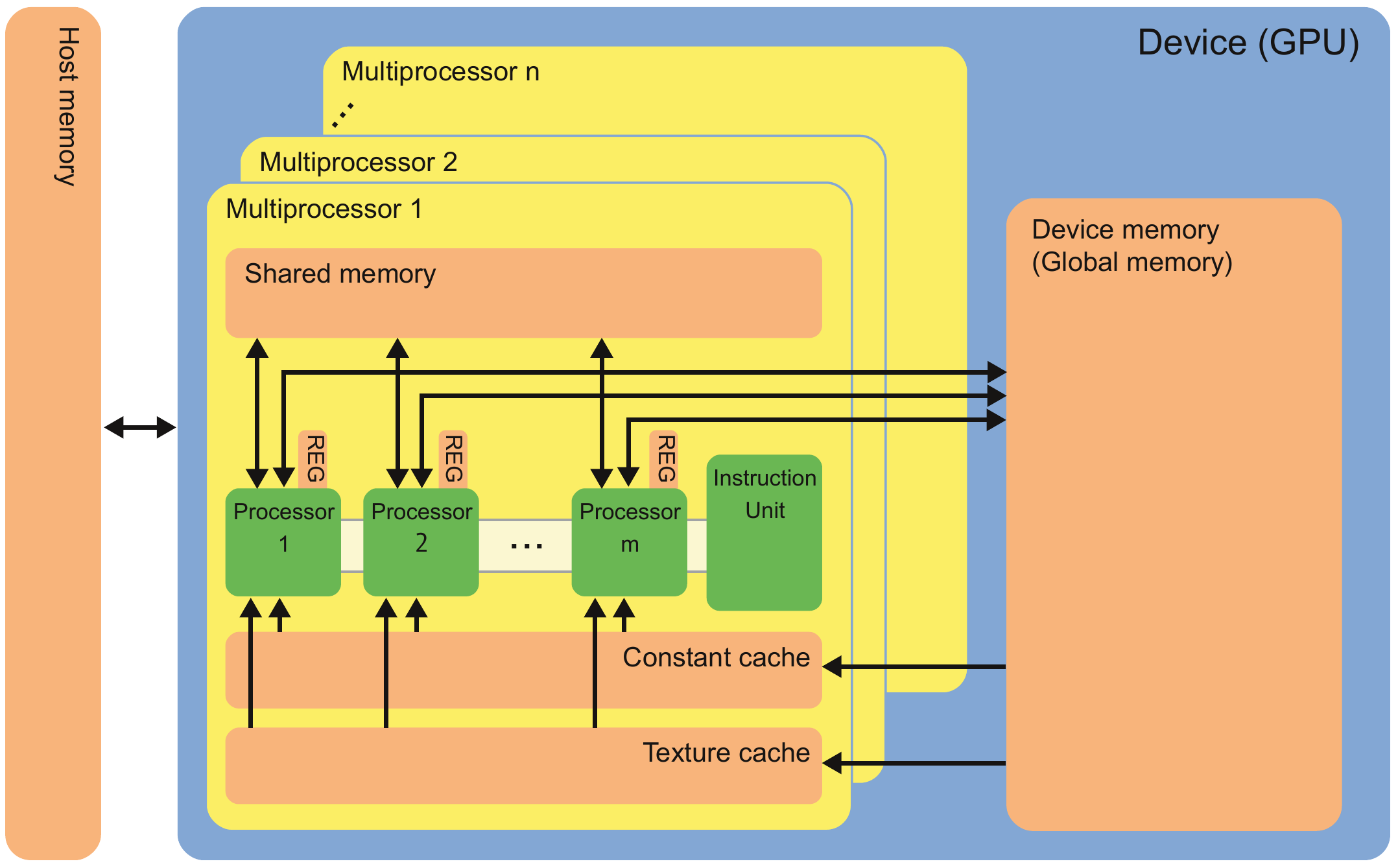}
  \caption{Diagrammatic representation of the hardware layout of recent NVIDIA GPUs.}
  \label{fig:hardware}
\end{figure}

Figure \ref{fig:hardware} shows a schematic representation of the NVIDIA GPUs used in
the work presented here. A GPU consists of a number of multiprocessors, each composed
of a number of single processing units which concurrently work with the same code on
different parts of a common data set. Of utmost importance to the efficient
performance of GPU programs is the organization of GPU memory, which comes in a
number of flavors:
\begin{itemize}
\item {\em Registers}: each multiprocessor is equipped with several thousand
  registers, access to which is local to each processing unit and extremely fast.
\item {\em Shared memory}: the processors combined in a multiprocessor have access to
  a small amount ($16$ KB for Tesla cards and $48$ KB for the Fermi architecture) of
  shared memory, which serves as a means of synchronization and communication between
  the threads in a block. This memory resides on-chip and can be accessed essentially
  without significant memory latency.
\item {\em Global memory}: this large amount of memory (currently up to $4$ GB) is on
  separate DRAM chips and can be accessed by each thread on each
  multiprocessor. Access suffers from a latency of several hundred clock cycles.
\item {\em Constant and texture memory}: these memory areas are of the same speed as
  global memory, but they are cached such that read access can be very fast. From
  device perspective they are essentially read-only.
\item {\em Host memory}: the memory of the host CPU unit cannot be accessed from
  inside GPU calculations. Memory transfers between global and device memory are
  important for communication with the ``outside world''.
\end{itemize}
Additionally, the recent Fermi architecture provides certain cache memories, but
since the previous Tesla architecture is used in the present work, I do not discuss
them here. In the CUDA framework, calculations are organized to match the layout of
the hardware: each multiprocessor executes (part of) a {\em block} of threads
concurrently, while the different blocks of a {\em grid} are assigned to separate
multiprocessors. To alleviate the large latency (in terms of clock cycles) of global
memory accesses, in an ideal setup there are many more threads in total than
available processors, such that a different (part of a) thread block can be scheduled
for execution while the threads of a given block wait for memory fetches or writes.

For maximum performance, implementations of scientific calculations have to take
these characteristics into account and, in particular, should ideally meet the
following design goals:
\begin{enumerate}
\item a large degree of locality of the calculations, reducing the need for
  communication between threads
\item a large coherence of calculations with a minimum occurrence of divergence of
  the execution paths of different threads
\item a total number of threads significantly exceeding the number of available
  processing units
\item a large overhead of arithmetic operations and shared memory accesses over global memory
  accesses
\end{enumerate}

\section{Double checkerboard Metropolis simulations}

\begin{figure}[tb]
  \centering
  \includegraphics[width=0.4\textwidth]{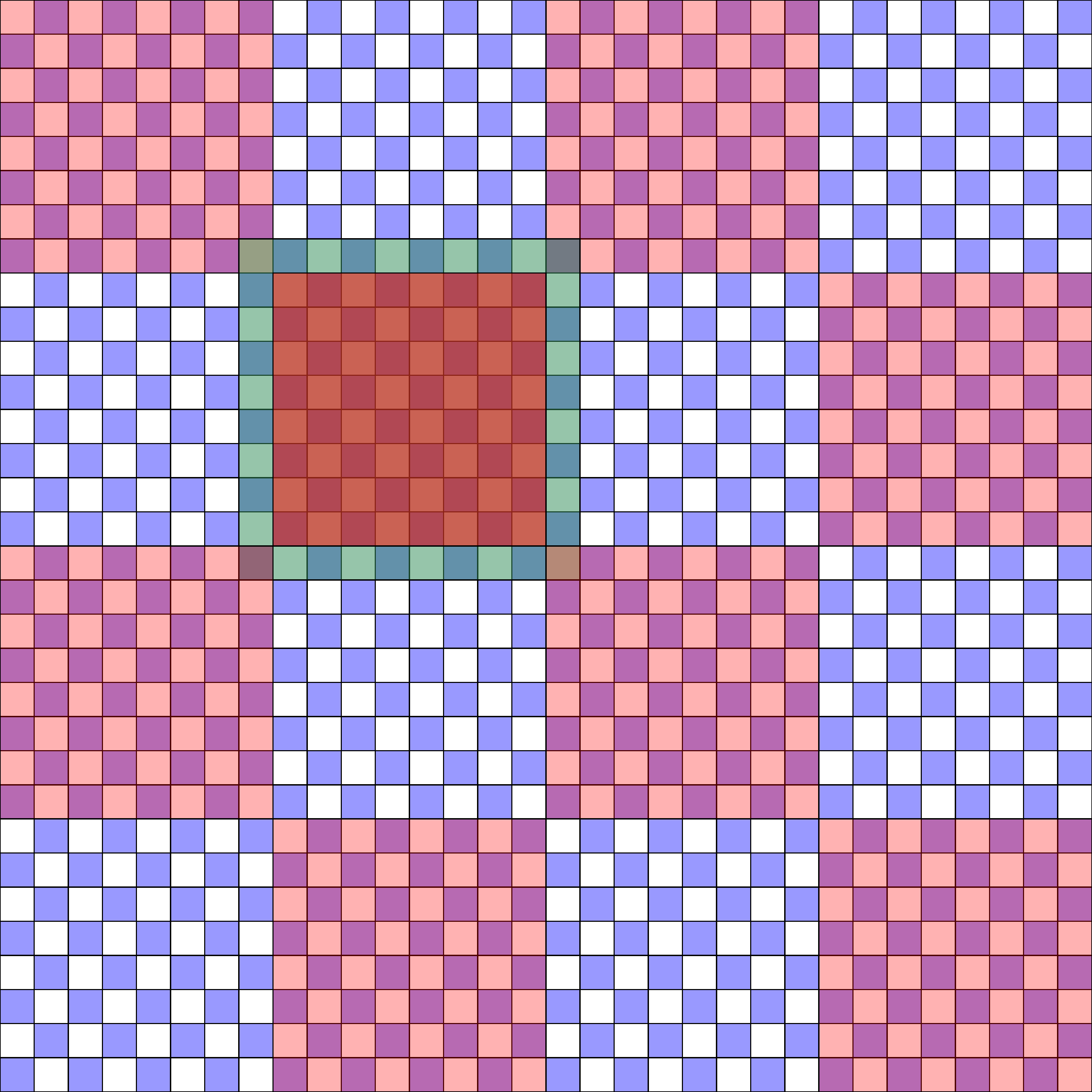}
  \caption{Double checkerboard decomposition of a $32^2$ square lattice for
    parallel Metropolis simulations on GPU. Each of the $B\times B = 4\times 4$ big
    tiles is assigned as a thread block to a multiprocessor, whose individual
    processors work on one of the two sub-lattices of all $T\times T = 8\times 8$
    sites of the tile in parallel.}
  \label{fig:checker}
\end{figure}

As a typical application in statistical physics, I studied the single-spin flip
Metropolis \cite{metropolis:53a} simulation of a nearest-neighbor, ferromagnetic
Ising model with Hamiltonian
\begin{equation}
  \label{eq:ising_hamiltonian}
  {\cal H} = -J\sum_{\l i,j\r}s_i s_j,\;\;\;s_i = \pm 1
\end{equation}
on square and simple cubic lattices of edge length $L$, using periodic boundary
conditions. A proposed flip of spin $s_i$ is accepted with the Metropolis probablity
\begin{equation}
  \label{eq:metropolis}
  p_\mathrm{acc}(s_i \mapsto -s_i) = \min\left[1,\,e^{-\beta\Delta E}\right],
\end{equation}
such that the updating decision can be drawn solely upon examining the states of spin
$s_i$ and its four (in 2D) resp.\ six (in 3D) neighbors. Hence, the necessary
calculations can be made local and highly parallel by using lattice decompositions of
the checkerboard type. The authors of Ref.~\cite{preis:09} used a single checkerboard
decomposition, working on strips or columns of the lattice. Since this setup does not
take the hierarchical memory organization into account, the spin field needs to
reside in global memory at all times, such that memory accesses are very
costly. Here, instead, I suggest to use a double checkerboard decomposition, whose
organization is in line with the hierarchic layout of GPU memory: for the
square-lattice system, on a first, ``coarse'' level, the lattice is divided into
$B\times B$ blocks. On a second, ``fine'' level, each block is decomposed, again in a
checkerboard fashion, into $T\times T$ sub-blocks. This is illustrated in
Fig.~\ref{fig:checker}. As a consequence of this decomposition, each large tile of
one of the two sub-lattices (``even'' and ``odd'') of the coarse decomposition can be
updated independently, and for each tile under consideration all sites of one
sub-lattice are again independent of each other. It is thus possible to load the
configuration of spins of one of the coarse tiles into shared memory, including an
extra surface layer of neighboring spins needed for calculating the local energy of
spins in the considered tile, cf.\ the shaded area in Fig.~\ref{fig:checker}. This
loading operation is distributed over the threads of a block, arranging memory
accesses to achieve coalescence \cite{cuda}.

In total, the simulation thus proceeds as follows:
\begin{enumerate}
\item A kernel is launched assigning all $B^2/2$ {\em even\/} tiles of the coarse
  checkerboard to a separate thread block, all of which are (depending on the number
  of multiprocessors available in hardware) executed in parallel.
\item The $T^2/2$ threads of each thread block cooperatively load the spin
  configuration of their tile plus a boundary layer into shared memory.
\item The threads of each block perform a Metropolis update of each {\em even\/}
  lattice site in their tile in parallel.
\item The threads of each block are synchronized, ensuring that all of them have
  completed the previous step.
\item The threads of each block perform a Metropolis update of each {\em odd\/}
  lattice site in their tile in parallel.
\item The threads of each block are again synchronized.
\item A second kernel is launched working on the $B^2/2$ {\em odd\/} tiles of the
coarse checkerboard in the same fashion as for the even tiles.
\end{enumerate}

In practice, the kernels for even and odd sub-lattices can be implemented as calls to
the same kernel, using an extra offset parameter to distinguish sub-lattices. To
leverage the effect of loading a tile's spin configuration into shared memory, a
generalized multi-hit technique \cite{berg:04} is employed for performing the
simulations, where steps $3$--$6$ above are repeated $k$ times. In this way, one
sub-lattice of the coarse checkerboard is updated several times before updating the
other sub-lattice. Close to criticality, the generalized multi-hit approach leads to
somewhat increased autocorrelation times \cite{weigel:10a}, which reduces the overall
efficiency of the implementation presented here in the vicinity of a critical
point. In view of the existence of efficient cluster algorithms for this case
\cite{kandel:91a}, however, single-spin flip Metropolis simulations are not the
algorithm of choice for this situation, anyway. The code for the Metropolis kernel
formulated here is extremely simple, taking up only around 60 lines (vs.\ around 300
lines in the implementation presented in Ref.~\cite{preis:09}). It can be downloaded
from the author's website \cite{weigel:gpu}.

\section{Results for the Ising model}

To actually perform the Metropolis updates, a stream of pseudo-random numbers is
required. It is clear that, for reasonable efficiency, each thread needs to have
access to an independent (sub-)stream of random numbers. For simplicity and the sake
of comparison, I here use an array of simple 32-bit linear congruential generators
(LCG) with identical multipliers, but randomly chosen initial seeds for each thread
\cite{preis:09}. It is clear that in view of the short period $p = 2^{32} \approx
10^9$ of the generators, most of the different sequences will have significant
overlap and, e.g., in a simulation with $10^7$ Monte Carlo sweeps of a $1024\times
1024$ system about $10^{13}$ random numbers are used, significantly exceeding the
period of the generator, and even more dramatically exceeding the value $\sqrt{p}$
considered to be safe when using LCGs \cite{knuth:vol2}. Somewhat surprisingly, for
the 2D model all simulation data are consistent with the exact results for the
internal energy and specific heat \cite{ferdinand:69a} with this setup. On the
contrary, when using an actually cleaner setup with {\em disjoint\/} sub-sequences of
the 32-bit LCG, and even when using disjoint sequences of an analogous 64-bit LCG
with period $p = 2^{64} \approx 10^{19}$, highly significant deviations are
encountered. For high-precision real-world applications, therefore, I suggest to use
different pseudo-random number generators, for instance of the Lagged Fibonacci type
\cite{brent:92}. The corresponding implementations will be discussed elsewhere
\cite{weigel:10a} .

\begin{figure}[tb]
  \centering
  \includegraphics[width=0.4\textwidth]{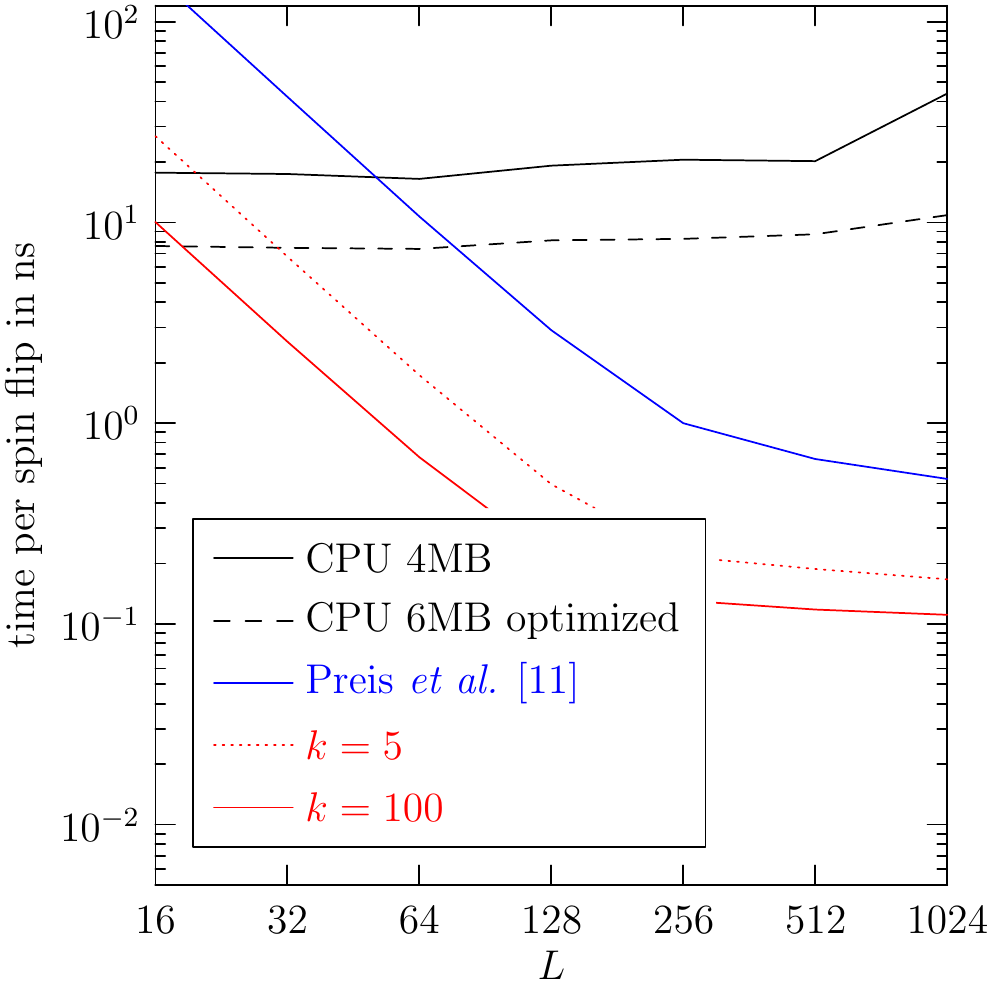}
  \caption{Computer times for a single spin flip of a Metropolis update simulation of
    the 2D Ising model on a square lattice of edge length $L$ using the double
    checkerboard decomposition and $k$-fold generalized multi-hit updates.  GPU times
    are for a Tesla C1060 device and CPU times for 3.0 GHz Intel Core 2 Quad
    processors with 4 MB and 6 MB of cache, respectively.}
  \label{fig:times}
\end{figure}

\begin{figure}[tb]
  \centering
  \includegraphics[width=0.4\textwidth]{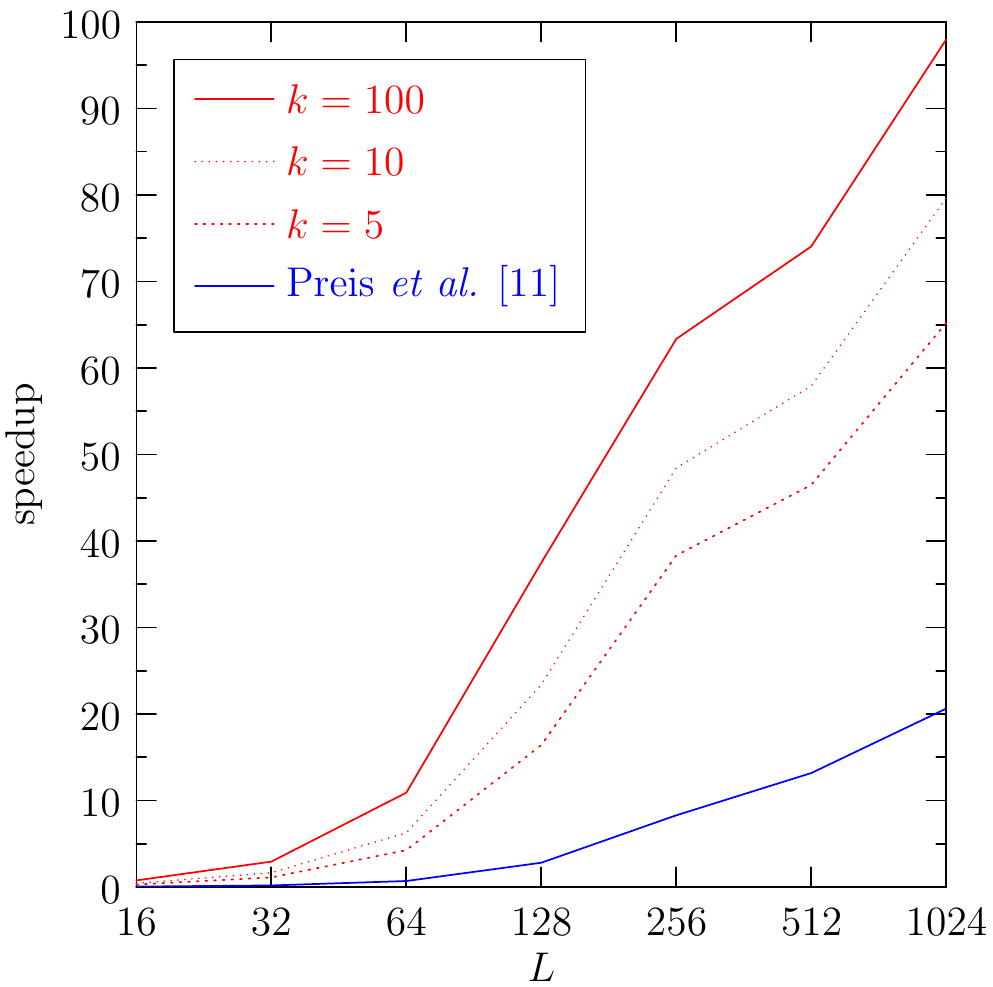}
  \caption{Speed-up factors of the double checkerboard GPU implementation for the 2D
    Ising model vs.\ the CPU code as a function of linear system size $L$.}
  \label{fig:speedup}
\end{figure}

For the 2D model, in Fig.~\ref{fig:times} the times for performing a single spin flip
are presented as a function of the linear system size $L$. The time required for the
measurement of elementary quantities such as the energy and magnetization is not
included in these figures, since pure spin-flip times over the years have developed
into a standard unit for comparing different architectures and implementations and
thus allow to compare to a host of previous calculations. GPU calculations have been
performed here on a Tesla C1060 device with 4 GB of RAM. By experimentation, for the
considered system sizes $16\le L\le 1024$, the optimal tile sizes are found to be $T
= 4$ for $L\le 64$, $T=8$ for $L=128$ and $T= 16$ for $128 < L\le 1024$. Using shared
memory and the multi-hit technique, single spin flip times down to about $0.1$ ns can
be achieved, significantly exceeding the performance reported in
Ref.~\cite{preis:09}. When comparing these results to CPU calculations, the question
arises whether multiple CPU cores should be taken into account \cite{dickson:10}. I
refrain her from doing so, and use serial CPU code as the {\em de facto\/} standard
of code used in most simulations on single CPUs. The CPU code used in
Ref.~\cite{preis:09} was a one-to-one copy of the GPU code. Just replacing it by code
more suitable for serial execution already results in a speed-up by a factor of
two. This observation, as well as the cache effect clearly visible in
Fig.~\ref{fig:times} as the size of the spin field of $4 L^2$ bytes reaches the size
of the cache, indicate that speed-up factors are a rather fragile measure of GPU vs.\
CPU performance. Trying a relatively fair comparison, using the somewhat optimized
code on a CPU with sufficiently large cache, results in the speed-up factors
presented in Fig.~\ref{fig:speedup}. Whereas compared to the CPU code used in
Ref.~\cite{preis:09} speed-ups of up to $400$ are observed, for the more realistic
comparison used here, a maximal speed-up of around $100$ is reached (vs.\ a speed-up
of around $20$ for the GPU code of Ref.~\cite{preis:09}). The double checkerboard
decomposition proceeds in a completely analogous way for the case of the 3D Ising
model, and in this case we achieve a maximum performance of around $0.24$ ns per
single spin flip with maximal speed-ups of almost $300$ compared to the corresponding
CPU code for a $256^3$ system (for this lattice size, the spin configuation is
significantly larger than the cache memory if using regular integer variables for the
spins).

It is obvious that the chosen problem and implementation come rather close to meeting
the design goals set out in Sec.~\ref{sec:hardware} and thus constitute a quite ideal
application. Indeed, we achieve a total throughput in excess of 100 GFLOP/s from the
chosen implementation which is at least of the same order of magnitude as the
theoretical peak performance of $933$ GFLOP/s for the Tesla C1060 card. The outlined
approach easily generalizes to simulations of more general spin models, in particular
models with continuous spins such as the Heisenberg model, where the large efficieny
of GPU devices with (single-precision) floating-point calculations comes into
play. For the case of disordered models, parallelism is also possible by working on
many disorder realizations concurrently. Combining such approaches with
(asynchronous) multi-spin coding, we achieve a performance of around $0.15$ ps per
single spin flip for the Edwards-Anderson Ising spin glass. These and further
extensions will be discussed in a separate publication \cite{weigel:10a}.

The author acknowledges support by the ``Center for Computational Sciences in Mainz''
(SRFN) as well as funding by the DFG through the Emmy Noether Programme under
contract No.\ WE4425/1-1.


\end{document}